\documentstyle[psfig]{mn}

\newif\ifAMStwofonts

\def\simgt{\hbox{\rlap{\raise 0.425ex\hbox{$>$}}\lower 0.65ex\hbox{$\sim$}}}
\def\simlt{\hbox{\rlap{\raise 0.425ex\hbox{$<$}}\lower 0.65ex\hbox{$\sim$}}}

\ifoldfss
  \ifCUPmtlplainloaded \else
    \NewTextAlphabet{textbfit} {cmbxti10} {}
    \NewTextAlphabet{textbfss} {cmssbx10} {}
    \NewMathAlphabet{mathbfit} {cmbxti10} {} 
    \NewMathAlphabet{mathbfss} {cmssbx10} {} 
  \fi
  \ifAMStwofonts
    \ifCUPmtlplainloaded \else
      \NewSymbolFont{upmath} {eurm10}
      \NewSymbolFont{AMSa} {msam10}
      \NewMathSymbol{\upi}     {0}{upmath}{19}
      \NewMathSymbol{\umu}     {0}{upmath}{16}
      \NewMathSymbol{\upartial}{0}{upmath}{40}
      \NewMathSymbol{\leqslant}{3}{AMSa}{36}
      \NewMathSymbol{\geqslant}{3}{AMSa}{3E}

      \let\leq=\leqslant 
       \let\ge=\geqslant
    \fi
  \fi
\fi 

\ifnfssone
  \newmathalphabet{\mathit}
  \addtoversion{normal}{\mathit}{cmr}{m}{it}
  \addtoversion{bold}{\mathit}{cmr}{bx}{it}
  \newmathalphabet{\mathbfit} 
  \addtoversion{normal}{\mathbfit}{cmr}{bx}{it}
  \addtoversion{bold}{\mathbfit}{cmr}{bx}{it}
  \newmathalphabet{\mathbfss} 
  \addtoversion{normal}{\mathbfss}{cmss}{bx}{n}
  \addtoversion{bold}{\mathbfss}{cmss}{bx}{n}
  \ifAMStwofonts
    \ifCUPmtlplainloaded \else
      \UseAMStwoboldmath
      \makeatletter
      \new@mathgroup\upmath@group
      \define@mathgroup\mv@normal\upmath@group{eur}{m}{n}
      \define@mathgroup\mv@bold\upmath@group{eur}{b}{n}
      \edef\UPM{\hexnumber\upmath@group}
      \new@mathgroup\amsa@group
      \define@mathgroup\mv@normal\amsa@group{msa}{m}{n}
      \define@mathgroup\mv@bold\amsa@group{msa}{m}{n}
      \edef\AMSa{\hexnumber\amsa@group}
      \makeatother
      \mathchardef\upi="0\UPM19
      \mathchardef\umu="0\UPM16
      \mathchardef\upartial="0\UPM40
      \mathchardef\leqslant="3\AMSa36
      \mathchardef\geqslant="3\AMSa3E

      \let\leq=\leqslant 
       \let\ge=\geqslant
    \fi
  \fi
\fi 

\ifnfsstwo
  \DeclareMathAlphabet{\mathbfit}{OT1}{cmr}{bx}{it}
  \SetMathAlphabet\mathbfit{bold}{OT1}{cmr}{bx}{it}
  \DeclareMathAlphabet{\mathbfss}{OT1}{cmss}{bx}{n}
  \SetMathAlphabet\mathbfss{bold}{OT1}{cmss}{bx}{n}
  \ifAMStwofonts
    \ifCUPmtlplainloaded \else
      \DeclareSymbolFont{UPM}{U}{eur}{m}{n}
      \SetSymbolFont{UPM}{bold}{U}{eur}{b}{n}
      \DeclareSymbolFont{AMSa}{U}{msa}{m}{n}
      \DeclareMathSymbol{\upi}{0}{UPM}{"19}
      \DeclareMathSymbol{\umu}{0}{UPM}{"16}
      \DeclareMathSymbol{\upartial}{0}{UPM}{"40}
      \DeclareMathSymbol{\leqslant}{3}{AMSa}{"36}
      \DeclareMathSymbol{\geqslant}{3}{AMSa}{"3E}

      \let\leq=\leqslant 
       \let\ge=\geqslant
    \fi
  \fi
\fi 

\ifCUPmtlplainloaded \else
  \ifAMStwofonts \else 
    \def\upi{\pi}
    \def\umu{\mu}
    \def\upartial{\partial}
  \fi
\fi
\title[]{The Space Density of low redshift AGN}
\author[D.Londish et al.]
	 {D.Londish$^{1}$, B.J.Boyle$^{1,2}$, D.J.Schade$^{3}$\\
$^1$ School of Physics, University of New South Wales, 
Kensington, NSW 2030, Australia\\
$^2$ Anglo-Australian Observatory, PO Box 296, Epping, NSW 1710, Australia\\
$^3$ Dominion Astrophysical Observatory, 5071 West Saanich 
Road, Victoria, V8X 4M6, Canada}

\begin{document}

\maketitle

\newcommand{\fmmm}[1]{\mbox{$#1$}}
\newcommand{\scnd}{\mbox{\fmmm{''}\hskip-0.3em .}}
\newcommand{\scnp}{\mbox{\fmmm{''}}}

\begin{abstract}

We present a new determination of the optical luminosity function
(OLF) of active galactic nuclei (AGN) at low redshifts ($z<0.15$)
based on Hubble Space Telescope (HST) observations of X-ray-selected
AGN.  The HST observations have allowed us to derive a true nuclear
luminosity function for these AGN. The resulting OLF illustrates a
two-power-law form similar to that derived for QSOs at higher
redshifts.  At bright magnitudes, $M_B<-20$, the OLF derived here
exhibits good agreement with that derived from the Hamburg/ESO QSO
survey.  However, the single power law form for the OLF derived from
the Hamburg/ESO survey is strongly ruled out by our data at $M_B>-20$.
Although the estimate of the OLF is best-fit by a power law slope at
$M_B<-20.5$ that is flatter than the slope of the OLF derived at
$z>0.35$, the binned estimate of the low redshift OLF is still
consistent with an extrapolation of the $z>0.35$ OLF based on pure
luminosity evolution.
\end{abstract}

\section{Introduction}

The QSO optical luminosity function (OLF) and its evolution with redshift has
been studied extensively for over three decades (see e.g. Schmidt
1968, Marshall et al.\ 1983, Boyle et al.\ 1988, 
Hewett, Foltz \& Chaffee 1993, La Franca \& Cristiani
1997).  This has led to a detailed picture of the QSO OLF over 
a wide range in redshift from $z\sim 0.3$ to $z>4$.  
In contrast, the local ($z<0.15$) QSO OLF is actually much more 
poorly determined, frustrating attempts to link QSO evolution
at moderate to high redshifts with nuclear activity in galaxies 
at the present epoch.

This is due to a number of factors associated with the compilation of
a suitable sample of local active galactic nuclei (AGN) with which to
derive the local OLF.  First, local AGN are relatively rare.  Their
space density is approximately 100 times less than that of normal
galaxies, and large area surveys are
required to yield a statistically useful sample. Secondly, many
selection techniques for local AGN suffer from morphological biases.
While surveys for stellar-like objects are clearly biased against
resolved AGN, galaxy-based surveys are equally biased against objects
with a dominant nuclear component.  Finally, accurate knowledge of
the nuclear OLF requires accurate subtraction of the light from the
host galaxy.  In the low luminosity AGN ($M_B>-23$) that constitute
the vast majority of the low redshift population, the light from the
host galaxy may dominate the nuclear luminosity. Even at relatively
low redshifts ($z \simeq 0.1$), seeing limitations imposed by ground
based observations limit accurate modelling of the luminosity profiles
of the central regions of AGN host galaxies to scales typically larger
than 1h$_{50}^{-1}$kpc.

A recent attempt to estimate the local AGN OLF has been carried out by
K\"ohler et al.\ (1997), hereinafter K97.  Using a sample of 27
candidates selected from the Hamburg/ESO objective prism survey, K97
derived a local AGN LF that exhibited a featureless power law form over
a wide range in absolute magnitudes $-24<M_B<-18$.  The form of the
low redshift AGN LF is thus very different from the two-power-law
luminosity function at higher redshifts ($z\ge0.5$). This is a
significant challenge for any theoretical model which seeks to connect
the evolution of QSOs at high redshift with the local AGN population.

The Hamburg/ESO survey covers an extensive area (611deg$^{2}$; now extended to 3700deg$^2$, see Wisotzki 2000) and is
free of morphological bias.  Unfortunately the spatial resolution
(1--2 arcsec) of the survey is not sufficiently good to permit an
accurate deconvolution of the galaxy and nuclear light even for the
lowest redshift AGN ($z<0.1$) in the sample. For the 0.07 $< z <$ 0.3
sample K97 used small-aperture, zero-point corrected $B$ band CCD magnitudes
which were subsequently  corrected to reflect nuclear luminosities by 
subtracting a template host galaxy value of $M_B = -21$ ;
for the AGN with $z<0.07$ corrections were calculated individually and 
ranged from 0.21 to 1.61 mag.

Until recently, AGN data sets studied with HST were either too small
or the samples on which they were based were too heterogeneous to
construct a reliable estimate of the local OLF.  We report here on the
estimate of the local nuclear OLF based on HST observations of 76 AGN
selected from a unbiased sample of X-ray selected AGN.  The sample is
part of the extensive Einstein Medium Sensitivity Survey (EMSS, Stocke
et al.\ 1991) which covers over 400deg$^{2}$ and has near-complete
($>96$ per cent) optical spectroscopic identification with no
morphological bias.  An earlier attempt to derive the low redshift OLF
using the EMSS was made by della Ceca et al.\ (1996).  They used 226
broad line AGN with $z<0.3$ to obtain a total (nuclear + host)
OLF. This OLF was then convolved with the observed distribution of
nuclear-to-total flux ratios for Seyfert 1 and 1.5 galaxies to yield a
nuclear OLF.  In this paper we propose to improve significantly on
this work by using the HST observtions to correct explicity for the
host galaxy light in each AGN.  In section 2 we report on the
measurement of the OLF from this sample, and in sections 3 and 4 we
present and discuss the results obtained, comparing them to the K97
results. We present our conclusions in section 5.

\section{Analysis}

\subsection{The data}

Details of the comprehensive HST imaging survey from which the sample
of AGN used in this analysis were drawn is presented in a paper by
Schade et al.\ (2000, hereinafter SBL). A full discussion of the methods
used to select and observe these AGN is presented by SBL,
thus only a brief description will be given here.

HST observations of 76 $z<0.15$ AGN selected from the EMSS survey
were carried out in the F814W ($I$) band, chosen to assist
in the detection of the redder host galaxy components over the bluer
nucleus.  These data were complemented by deeper ground-based observations
in the $B$ and $R$ bands for 69 AGN in the survey. A
simultaneous three-component parametric model fit to the $B$, $R$ and
$I$ images was performed for each AGN in the sample to derive
magnitudes for the nuclear point source, bulge and disk components in each 
object. Despite the improved spatial resolution afforded with the
HST, the fitting procedure is complex and error estimation required
significant modelling of the fitting process. For host-dominated
objects uncertainties in $M_B\rm{(host)}$\footnote{In the SBL study
all magnitudes were based on the AB system.  Nuclear $B{\rm (AB)}$
magnitudes were derived by applying a mean $(B-I)_{\rm (AB)}=0.2$ 
colour correction to the nuclear $I{\rm (AB)}$-band magnitudes
obtained from the fit to the HST data.  For objects of this colour,
there is a negligible colour term between 
$B$ and $B{\rm (AB)}$ passbands. For the purposes of this analysis we
have therefore assumed  $M_B{\rm (AB)}=M_B$ for the nuclear regions.}  
were typically $\pm0.25\,$
magnitudes in the region of the $M_B,z$ plane where the AGN are found, but
increased to $\pm 0.5\,$mag where the nucleus was dominant.
Similarly, errors in nuclear magnitudes were $\pm$0.25 mag for bright
nuclei but as much as $\pm 0.5$ mag for host-dominated objects.

In total, nuclear $M_{B}$ magnitudes were obtained for 66 AGN
in the sample (10 had no detectable nuclear component), and these data
form the basis for the calculation of the OLF below.   
Nuclear absolute magnitudes were found to lie in the range $-14.6 >
M_B > -24.1$. The region of the AGN $M_B{\rm(nuc)}, z$ plane
sampled in this study is shown in Fig.\ 1. 

Since we are attempting to contruct an OLF from an X-ray-selected
sample we need to ensure that there is a good correlation between
nuclear optical and X-ray luminosity for objects in the SBL sample.
We used the monochromatic $2\,$keV X-ray luminosity, $L_{\rm
2keV}$(nuc) and 2500\AA\ UV fluxes, $L_{\rm 2500A}$(nuc), listed in
the SBL paper.  $L_{\rm 2500}$\AA(nuc) is based on the fitted nuclear
$M_B{\rm(nuc)}$, assuming a power-law optical/UV spectrum of the form
$f_\nu\propto\nu^{-0.5}$.  Errors on the X-ray flux range from 5 per
cent for the brightest X-ray sources to 25 per cent for the faintest
X-ray sources (Gioia 1990).  Although the X-ray luminosity for each
source is expected to be dominated by the AGN, it is impossible to
rule out some contribution from the host galaxy, particularly for the
lowest luminosity sources.

The least-squares fit to the observed relationship between $L_{\rm
2keV}$(nuc) and $L_{\rm 2500A}$(nuc) plotted in Fig.\ 2 gives a relation of
the form $L_{\rm 2keV}{\rm(nuc)} \propto L_{\rm 2500A}{\rm(nuc)}^{0.82 \pm 0.08}$, consistent with other studies (e.g.\ Green et al.\ 1995).

\begin{figure}
\centerline{\psfig{file=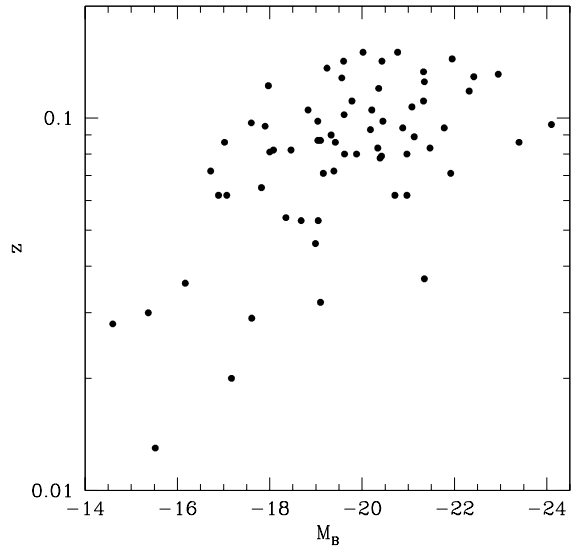,width=3.2in}}
\caption{\footnotesize Nuclear absolute magnitude {\it v.} redshift for 
the 66 AGN with point source detections in SBL.}
\end{figure}

\begin{figure}
{\centerline{\psfig{file=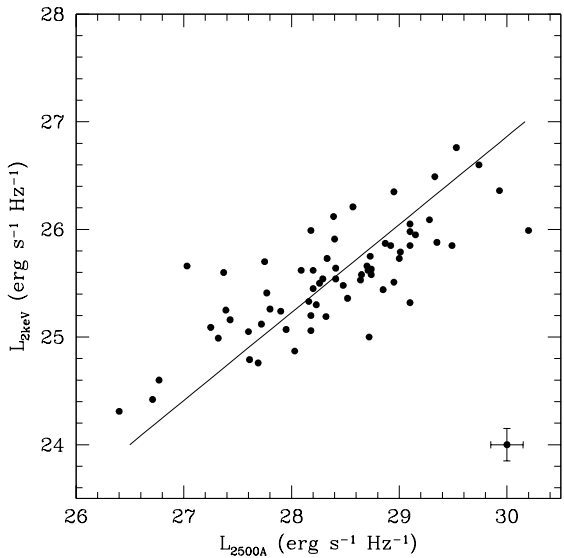,width=3.2in}}}
\caption{\footnotesize The correlation between nuclear UV and X-ray flux 
from the central component. The line is the slope of best fit calculated 
as $0.82 \pm 0.08$. Typical error on the flux estimates are shown  
in the lower right hand corner.}
\end{figure}

\subsection{The $1/V_a$ OLF estimate}

Space densities were derived using the $1/V_a$ estimator (Avni \&
Bahcall 1980).  The OLF was constructed from the summed contributions
of $n$ AGN using:

\begin{displaymath}
\Phi (M_B), z) = \sum_{i=1}^n \frac{1}{V_{a}^i} \  \delta (M_B^i-M_B)
\end{displaymath}
$V_a^i$ being the accessible co-moving volume of the $i^{\rm th}$ AGN.  The
estimate of $V_a$ was based on $z_{\rm max}$ derived from the
AGN's X-ray flux and EMSS flux limits, assuming an X-ray spectral
index $\alpha_{\rm X}=1$; $f_{\nu} \propto \nu^{-\alpha_{\rm X}}$. To construct an
optical OLF, we binned the $1/V_a$ estimates according to their optical nuclear
$M_{B}$ magnitudes. We computed Poisson errors on the binned
estimates of the OLF using $\sigma = \left(\sum \frac{1}{(V_a^i)^2}
\right) ^{0.5}$.\\

Not all 127 AGN with $z < 0.15$ in the EMSS were observed by SBL
and so a straightforward normalising factor of 0.6 (76/127) was applied to
the area coverage function when computing the accessible volume. KS
tests confirmed that the redshift and flux distribution for the SBL 
sample is consistent with the sample being drawn at random from the 
$z < 0.15$ EMSS parent sample.

The OLF was calculated at 1-mag intervals for the full redshift range
$z \leq 0.15$.  We made no correction for evolution across the
redshift bin. We also constructed separate OLFs for AGN in elliptical
and spiral hosts to investigate any host-related trends.

\section{Results}

The differential OLF calculated for an Einstein-de Sitter universe in
which $H_0 = 50 h_{50}$ km s$^{-1}$ Mpc$^{-1}$, $\Omega_{\rm M}=1$,
$\Omega_\Lambda=0$, is presented in Table 1 and plotted in Fig. 3, 
over plotted with data from K97 (their Table 5). As a comparison values
using total galaxy luminosity (host + nucleus) are shown in Table 2 and
 plotted in Figure 5.

\begin{table}
\begin{center}
\caption{Nuclear Optical Luminosity Function}
\begin{tabular}{|c|c|c|}
\hline
\multicolumn{1}{|c|}{$M_B$ (nucleus)} & 
\multicolumn{1}{|c|}{$\Phi$ (Mpc$^{-3}$ mag$^{-1}$)} &
\multicolumn{1}{|c|}{Objects}\\
\hline
$-$24.00  &  $4.30 \pm    4.30 \times 10^{-8}$  &    1\\
$-$23.00  &  $3.98 \pm    3.98 \times 10^{-8}$  &    1\\
$-$22.00  &  $1.31 \pm   7.59 \times 10^{-8}$  &    3\\
$-$21.00  &  $5.61 \pm    1.91 \times 10^{-7}$  &   10\\
$-$20.00  &  $1.43 \pm    5.69 \times 10^{-7}$  &   14\\
$-$19.00  &  $1.90 \pm    5.63 \times 10^{-7}$  &   16\\
$-$18.00  &  $1.57 \pm    7.40 \times 10^{-7}$  &    7\\
$-$17.00  &  $3.42 \pm    1.63 \times 10^{-6}$  &    8\\
$-$16.00  &  $7.80 \pm    5.18 \times 10^{-7}$  &    3\\
$-$15.00  &  $3.38 \pm    2.54 \times 10^{-6}$  &    2\\
$-$14.00  &  $3.36 \pm    3.36 \times 10^{-6}$  &    1\\

\hline

\end{tabular}
\end{center}
\end{table}

\begin{table}
\begin{center}
\caption{Total Optical Luminosity Function}
\begin{tabular}{|c|c|c|}
\hline
\multicolumn{1}{|c|}{$M_B$ (host + AGN)} & 
\multicolumn{1}{|c|}{$\Phi$ (Mpc$^{-3}$ mag$^{-1}$)} &
\multicolumn{1}{|c|}{Objects}\\
\hline
$-$24.00  &  $4.30 \pm    4.30 \times 10^{-8}$  &    1\\
$-$23.00  &  $1.19 \pm    0.69 \times 10^{-8}$  &    3\\
$-$22.00  &  $9.92 \pm    2.55 \times 10^{-7}$  &    17\\
$-$21.00  &  $4.91 \pm    1.31 \times 10^{-6}$  &   28\\
$-$20.00  &  $6.98 \pm    3.69 \times 10^{-6}$  &   14\\
$-$19.00  &  $2.48 \pm    2.30 \times 10^{-6}$  &   2\\
$-$18.00  &  $1.09 \pm    1.09 \times 10^{-6}$  &    1\\

\hline

\end{tabular}
\end{center}
\end{table}

The K97 data comprises 27 objects extending out to a redshift
of 0.3.  Of these, eight were at redshifts greater than the $z = 0.15$
cut-off adopted in the SBL sample, all of which have $M_B < -24$, i.e. 
brighter than the most luminous AGN in the SBL sample.

In Fig.\ 3 we have also plotted two predictions of the $z<0.15$ OLF
based on the luminosity evolution models of Boyle et al.\ (2000).
These authors fit a variety of evolutionary models to a data set
comprising over 6000 QSOs with $M_B<-23$ and $0.35<z<2.3$ selected
from the 2dF QSO redshift survey (Boyle et al.\ 1999) and the Large
Bright QSO survey (LBQS, Hewett et al.\ 1995).  Boyle et al.\ (2000)
found that luminosity evolution models provided acceptable fits to the
data, with exponential evolution ($L^*\propto e^{k\tau}$) as a
function of look back time ($\tau$) favoured for a $q_0=0.05$ universe
and as a general second order polynomial with redshift ($L^*\propto
10^{k_1z+k_2z^2}$) for $q_0=0.5$.  The extrapolated $z<0.15$ OLFs for
the best-fitting `exponential' and `polynomial' models are shown as
the short- and the long-dashed lines respectively in Fig.\ 3.  The
model OLFs have been plotted over the magnitude range consistent with
the corresponding range (with respect to $M_B^*$) over which they were
derived at $z>0.35$.  We obtained a reduced $\chi^2=1.0$ for the exponential
model fit to the SBL data at $M_B <-19$, but were able to reject the
extrapolation of the polynomial model at the 99 per cent confidence level.

\begin{figure}
{\hspace*{0.1in}{\psfig{file=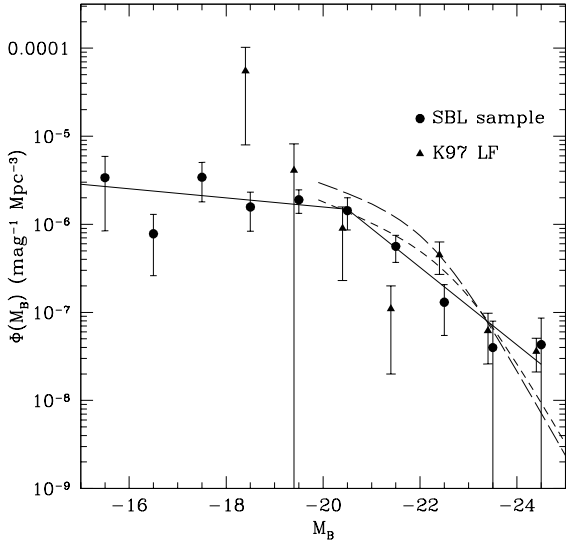,height=3.2in}}}
\caption{\footnotesize Binned OLF of the 66 AGN in the SBL survey
(filled circles) compared with data from K97 (triangles). The solid
line denotes the least squares fit to the data. Predicted $z<0.15$
OLFs based on luminosity evolution models of Boyle et al.\ (2000) are
also shown (short dashed line: `exponential' model, long dashed
line: `polynomial' evolution, see text for details).}
\end{figure}

\section{DISCUSSION}

There is good agreement both in slope and normalisation between our
estimate of the OLF and the K97 OLF at $M_B<-20$.  However, at fainter
magnitudes the two estimates diverge.  Our estimate of the OLF turns
over to a much flatter slope whereas the K97 OLF continues to rise
steeply.  However, there are only three AGN in the K97 sample with
$M_B>-20$, whereas the SBL sample contains 37 AGN at these fainter
magnitudes.  It is therefore most probable that the difference between
the two data-sets ($2\sigma$) at these magnitudes is simply due to
small number statistics in the K97 sample.

It is possible that the X-ray selection used to generate the SBL
dataset is systematically biased against AGN with low optical
luminosity.  However, the correlation between $L_{\rm 2keV}{\rm(nuc)}$
and $L_{\rm 2500A}{\rm(nuc)}$ plotted in Fig.\ 2 demonstrates that
there is no systematic trend for objects with lower optical
luminosities to exhibit relatively weaker X-ray-to-optical flux
ratios.  Indeed, the derived relation $L_{\rm 2keV}{\rm(nuc)} \propto
L_{\rm 2500A}{\rm(nuc)}^{0.82\pm0.08}$ implies the reverse, i.e. that AGN with
lower optical luminosities have stronger X-ray-to-optical flux ratios.

In common with other groups (La Franca \& Cristiani 1997, 
Goldschmidt \& Miller 1998), K97 have used their determination of
the low redshift OLF to claim that the slope of the bright end
of the OLF ($\Phi (L)$) flattens significantly from $\Phi(L)\propto
L^{-3.6}$ at $z>0.6$ to $\Phi(L)\propto L^{-2.5}$ at $z<0.3$.
Such an observation would rule out pure luminosity evolution models,
in which the shape of the OLF remains invariant with redshift.
This is in marked constrast with our result that an extrapolation of 
the exponential form of a pure luminosity evolution model derived at
$z>0.35$ still provides an adequate fit to the $z<0.15$ OLF.

To investigate the discrepancy between these results,
we fitted our binned estimate of the $z<0.15$ OLF with a two-power-law
model of the form:
$$\Phi(L)\propto L^{\alpha}\qquad L>L^*$$
$$\Phi(L)\propto L^{\beta}\qquad L<L^*$$

\noindent
Fixing a `break' luminosity at $L^*
\equiv M_B^*=-20.5$, we derived slopes of $\alpha=-2.1\pm 0.3$ and
$\beta=-1.1\pm 0.1$ using a weighted least squares technique. 
This fit is over-plotted as the solid line on
Fig.\ 3.  This slope for $\alpha$ is indeed flatter than that derived
at high redshift and is consistent with other estimates of the slope
of the low redshift OLF, including the most recent determination of
the bright end slope of the X-ray QSO LF ($\alpha=-2.6$) by Miyaji et
al.\ (1998). However, the value of $\alpha$ derived in this crude
fashion is strongly dependent the choice of $M_B^*$, and the inclusion
of the two brightest bins that each contain a single object.  By
choosing a `break' magnitude of $M_B=-21.5$, and restricting
consideration of the data points in the OLF to those bins which
contain more than one object, we can increase this value to $\alpha =
-2.6\pm0.3$.  This is, admittedly, a very crude analysis and more
sophisticated fitting of a smooth two-power-law function similar to
that used to fit the higher redshift OLF would yield a more accurate
estimate of the statistical errors associated with fitting the OLF.

However, it is also likely that systematic errors play an equally
important role in the determination of the local OLF.  We attempted to
estimate the sizes of such errors by first exploring the factor used to
correct for the host galaxy luminosity. 
\begin{figure}
{\hspace*{0.1in}{\psfig{file=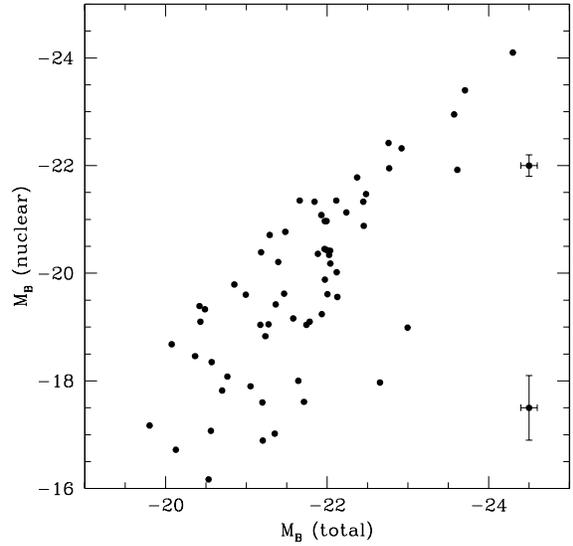,height=3.2in}}}
\caption{\footnotesize The relation between nuclear and total galaxy
luminosity for the SBL dataset. At fainter magnitudes the distribution
is very broad, with no clear correlation between nuclear and total
luminosity. Typical errors for the nuclear component are indicated;
errors are greater at fainter nuclear magnitudes.}
\end{figure}

In the SBL sample the host galaxy luminosity was explicitly removed
using fits to the individual HST images.  For the bulk of their sample
(i.e. 0.07 $ < z < 0.3$, or $M_B < -22$) K97 relied on a two-step
procedure using corrected CCD magnitudes measured in an aperture of
diameter approximately equal to that of the seeing disk, hence
replacing total magnitudes with small aperture magnitudes.
Subsequently, a further correction factor for host galaxy luminosity
was applied by adopting a template host galaxy of $M_B = -21$.

In Fig.\ 4 we have plotted nuclear luminosity against the total
galaxy luminosity for the SBL sample.  As also found by della Ceca et
al.\ (1996), although the more powerful AGN reside in the more luminous
hosts, the distribution of the ratio between nuclear and total
luminosity is not constant; indeed the spread becomes very large at
total absolute magnitudes fainter than $\sim M_B{\rm(nuc)} =-22.5$.

We re-computed our estimate of the $z<0.15$ OLF based on the SBL
dataset using total, instead of nuclear, absolute magnitudes.  The
resulting OLF is shown in Fig.\ 5, with the both the original fit
derived for the OLF and the extrapolated model fits plotted as a
comparison.  We find that although the bright end of the OLF has
steepened appreciably to $\alpha \sim -3$, both model fits are now
clearly incompatible with the OLF computed in this fashion.

The treatment of galaxy luminosities can thus result in significant
differences to the estimate of the OLF at zero redshift.  Note also
that the evolutionary models have been derived from high redshift OLFs 
uncorrected for host galaxy light.  Although the assumption
that the host galaxy light makes an increasingly small contribution to
the total luminosity of QSOs at high redshift may well be correct,
some spectacular counter-examples have already been discovered
(Aretxaga et al.\ 1995, Brotherton et al.\ 1999).

We conclude that the large statisitical and systematic errors
associated with the determination of the low redshift OLF and
extrapolation of the OLF at higher redshifts make it difficult to rule
out luminosity evolution models on the basis of shape of the low
redshift OLF.

\begin{figure}
{\hspace*{0.1in}{\psfig{file=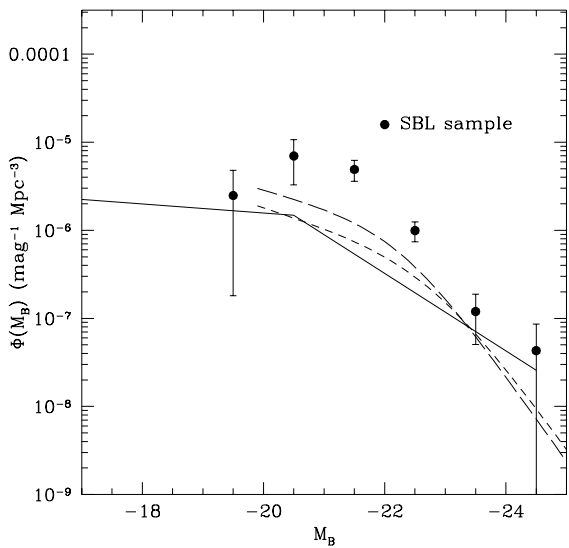,height=3.2in}}}
\caption{\footnotesize Binned LF of the 66 AGN in the SBL using total
absolute magnitudes.
 Solid line: least-squares fit to original OLF;  Short dashes
line: OLF predicted by exponential evolution model; long dashes: OLF
predicted by polynomial evolution model, see text for details.}
\end{figure}

We also attempted to derive nuclear OLFs for AGN with bulge-dominant
(E/S0) and disk-dominant (spiral) hosts.  Following SBL, the
distinction between the two broad classes and elliptical was made on
the basis of the bulge-to-total light ratio, $B/T$, for the host
galaxy.  Galaxies with $B/T>0.5$ were classified as E/S0 (44 in the
sample), those with $B/T \leq 0.5$ as spiral (22 in the sample).  The
nuclear OLFs for different types of host galaxy is plotted in Fig.\ 6
(after re-normalising the E/SO LF to the same space density as the
Sa/Sb LF).  There are no significant differences between the two OLFs;
confirmed by a KS test on the cumulative luminosity distributions for
spiral and E/S0 hosts.

\begin{figure}
{\hspace*{0.2in}{\psfig{file=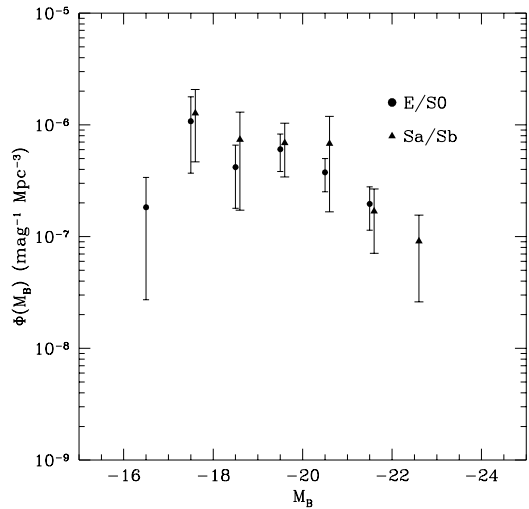,height=7.5cm}}}
\caption{OLFs for the SBL sample split according to host galaxy. 
The points for the E/SO hosts are
displaced to fainter magnitudes by $0.1\,$mag for clarity.}
\end{figure}

This result is consistent with observation by SBL that $B/T$ for the
host galaxy was independent of nuclear luminosity.  In contrast, we
note that McLure et al.\ (1999) found that the fraction of elliptical
hosts increased significantly amongst the highest luminosity
AGN. However, these observations are still based on relatively small
datasets and the McLure et al.\ (1999) study predominantly samples a
different luminosity regime ($M_B<-23$) to that under investigation
in this analysis.

\section{Conclusions}
Results from a comprehensive, unbiased X-ray selected sample of AGN
using the 0.1 arcsec resolving power of the HST, have enabled the first
direct estimate of the nuclear OLF for AGN to be constructed.  

The OLF derived illustrates a two power law form similar to that
derived for QSOs at higher redshifts and as such is different to the
largely featureless power-law OLF claimed for the low redshift
AGN identified in the Hamburg/ESO QSO survey.  However, any
discrepancy only occurs at $M_B>-20$, where previous estimates of the
OLF from the Hamburg/ESO survey are dominated by statistical errors
arising from small number statistics.

The OLF is consistent with an extrapolation of the exponential pure
luminosity evolution derived at $z>0.35$ by Boyle et al.\ (2000),
although the 'best-fit' slope for the bright-end slope of the OLF is
flatter than predicted by such pure luminosity evolution models.

Given the large uncertainties associated with current estimates of the
low redshift OLF (not least in the present analysis) and the
extrapolation of evolutionary models to low redshift, it is almost
certainly premature to rule out luminosity evolution on the basis of
the current determinations of the low redshift OLF.

Further detailed imaging work on optically-selected samples of
low-moderate redshift AGN/QSOs will clearly help resolve this issue.
With the superior imaging capability of the new generation of
ground-based telescopes (Keck, Gemini) we may look forward to such
data being obtained in the near future.

\section{Acknowledgements}
We thank the referee, Lutz Wisotzki, for a number of useful suggestions which
significantly improved the presentation and discussion of the results
in this paper.


\begin{thebibliography}{}

\bibitem[Aretxaga et al. 1980]{a95}Aretxaga I., Boyle B.J., Terlevich R.J., 1995, MNRAS, 275, L27 
\bibitem[Avni and Bahcall 1980]{ab}Avni Y., Bahcall J.N., 1980, ApJ 235, 717
\bibitem[Boyle et al. 1988]{bsp}Boyle B.J., Shanks T., Peterson B.A., 1988, MNRAS, 235, 935
\bibitem[Boyle et al 1999]{b99}Boyle B.J., Croom S.M., Smith R.J., Shanks T., Miller L., 
Loaring N., 1999, Phil.\ Trans.\ R.\ Soc.\ Lond.\ A., 357, 185
\bibitem[Boyle et al 2000]{b00}Boyle B.J., Shanks T., Croom S.M., 
Smith R.J., Miller L., Loaring N., 2000, MNRAS, submitted
\bibitem[Brotherton et al. 1999]{br99}Brotherton M.\ et al., 1999, ApJ, 520, 87
\bibitem[della Ceca et al. 1996]{d96}della Ceca R., Zamorani G., Maccacaro T., Setti G., Wolter A., 1996, ApJ 465, 650
\bibitem[Gioia et al. 1990]{g90}Gioia et al. 1990, ApJS 72, 56
\bibitem[Goldschmidt \& Miller 1998]{gm98}Goldschmidt P., Miller L., 1998, MNRAS, 293, 107
\bibitem[Green et al 1995]{g95}Green P.J. et al. 1995, ApJ 450, 51
\bibitem[Hewett et al. 1993]{hfc93}Hewett P.C., Foltz C.B., Chaffee F.H., 1993, ApJ, 406, L43
\bibitem[Hewett et al. 1995]{hfc95}Hewett P.C., Foltz C.B., Chaffee F.H., 1995, AJ, 109, 1498
\bibitem[Koehler et al. 1997]{kgrw97}K\"ohler T., Groote D., Reimers D., Wisotzki L., 1997, AA, 325, 502
\bibitem[La Franca \& Cristiani 1997]{lc97}La Franca F., Cristiani S., 1997, AJ, 113, 1517 
\bibitem[Marshall et al. 1983]{mtaz83}Marshall H.L., Tananbaum H., Avni Y., Zamorani G., 1983, ApJ, 269, 35
\bibitem[McLure et al. 1999]{m99}McLure R.J., Dunlop J.S., Kukula M.J., Baum S.A., O'Dea C.P., Hughes D.H., 1999, MNRAS, 308, 377
\bibitem[Miyaji et al. 1998]{m98}Miyaji T., Hasinger G., Schmidt M., 1998, astro-ph/9809398
\bibitem[Schade et al. 2000]{sbl}Schade D., Boyle B.J., Letawsky M., 
2000, MNRAS, in press (astro-ph/9912294)
\bibitem[Schmidt 1968]{s68}Schmidt M., 1968, ApJ, 151, 393
\bibitem[Stocke et al. 1991]{s91}Stocke J.T. et al., 1991, ApJS 72, 567
\bibitem[Wisotzki 2000]{w00}Wisotzki L., 2000, A\&A 253, 853
\end{thebibliography}
\end{document}